\documentstyle[twocolumn,graphicx,aps]{revtex}

\begin{document}

\draft

\title{Phonon spectrum and dynamical stability of a quantum
degenerate Bose-Fermi mixture}

\title{Phonon spectrum and dynamical stability of a quantum degenerate
Bose-Fermi mixture}

\author{Han Pu,$^1$ Weiping Zhang,$^1$ Martin Wilkens,$^2$ and Pierre
Meystre$^1$} 
\address{
{$^1$Optical Sciences Center, The University of Arizona,
Tucson, AZ 85721}\\
{$^2$ Universit\"{a}t Potsdam, Institut f\"{u}r Physik, Am Neuen Palais 10,
14469 Potsdam, Germany} \\ (\today)
\\ \medskip}\author{\small\parbox{14.2cm}{\small\hspace*{3mm}
We calculate the phonon excitation spectrum in a zero-temperature
boson-fermion mixture. We show how the sound velocity changes due to the
boson-fermion interaction and we determine the dynamical stability regime of a
homogeneous mixture. We identify a resonant phonon-exchange interaction
between the fermions as the physical mechanism leading to the instability.\\
\\[3pt]PACS numbers: 03.75.Fi, 05.30.Fk, 67.40.Db}} \maketitle

\narrowtext
New developments in the field of ultracold atoms are creating
opportunities for detailed studies of quantum degenerate Fermi
gases\cite{jin,hulet}. The Pauli exclusion principle results in
the lack of $s$-wave scattering and the suppression of higher
order scattering processes at low temperatures, so that the direct
evaporative cooling spin-polarized fermionic atoms does not work.
To circumvent this obstacle, one may either cool a Fermi gas
containing two different spin states\cite{jin} or mix the fermions
with a gas of bosonic atoms and sympathetically cool this
mixture\cite{hulet}. Both methods have been successfully
implemented experimentally to create degenerate Fermi gases. The
experimental progress along the latter direction has stimulated
considerable renewed interest in studying Bose-Fermi mixtures in
the quantum degenerate regime\cite{molmer,tosi,stoof,smith}.

The study of ultracold Bose-Fermi mixtures dates back to the
1960s, when the discovery that $^3$He is miscible in $^4$He
prompted the study of dilute solutions of $^3$He in superfluid
$^4$He\cite{baym}. In that system, however, the physical picture
is complicated by the strong interatomic interactions. As such,
weakly-interacting alkali atomic vapor quantum gases provide us
much cleaner systems to study the properties of Bose-Fermi
mixtures. In this paper, we analyze the spectrum of bosonic
elementary excitations in a Bose-Fermi mixture at zero
temperature. In the case of an homogeneous system, we identify the
stability regime of the mixture and calculate the sound velocity
of the bosonic phonons under the influence of the fermions. We
identify the physical mechanism leading to the instability of the
mixture as a resonant phonon-exchange coupling between fermions
below and above the Fermi sea.

Our starting point is the Hamiltonian describing the interaction
between a bosonic and spin-polarized fermionic system
\begin{eqnarray}
\hat{H} &&= \int d^3 r\,\hat{\psi}_B^\dagger \left(\hat{H}_B^{(0)}-\mu_B+
\frac{g}{2} \hat{\psi}_B^\dagger \hat{\psi}_B \right) \hat{\psi}_B \nonumber \\
&&\;+ \int d^3r\, \hat{\psi}_F^\dagger \left(\hat{H}_F^{(0)}-\mu_F \right) 
\hat{\psi}_F   + f\int
d^3r\,\hat{\psi}_B^\dagger \hat{\psi}_B \hat{\psi}_F^\dagger \hat{\psi}_F ,
\label{h}
\end{eqnarray}
where $\hat{H}_{\alpha}^{(0)} \equiv
\hat{T}_{\alpha}+V_{\alpha}$ ($\alpha = B,\;F$) and
the subscripts $B$ and $F$ represent `boson' and `fermion',
respectively; $\hat{T}_\alpha$, $V_\alpha$ and $\mu_\alpha$ are
the associated kinetic energy, trapping potential, and chemical
potential, and $g$ and $f$ represent the boson-boson and
boson-fermion interaction strengths. In terms of the $s$-wave
scattering lengths $a$, they are given by
\begin{eqnarray*}
g = 4\pi \hbar^2a_B/m_B,\;\;
f = 2\pi \hbar^2 a_{BF}/m_r,
\end{eqnarray*}
with $m_r=m_B m_F/(m_B+m_F)$ being the reduced mass. All higher order
scattering processes are neglected.

To determine the excitation spectrum of the Bose condensate, we
decompose the field operators as
\begin{eqnarray}
\hat{\psi}_B ({\bf r},t) &=& \phi_B({\bf r}) + \hat{\xi}_B({\bf
r},t) ,\nonumber\\
\hat{\psi}_F^\dag ({\bf r},t) \hat{\psi}_F ({\bf
r},t)&=&  n_F({\bf r})+\delta \hat{\rho}_F({\bf r},t),
\label{decom}
\end{eqnarray}
where $\phi_B({\bf r}) = \langle \hat{\psi}_B ({\bf r},t) \rangle$
is the condensate wave function and  $n_F({\bf r}) = \langle
\hat{\psi}_F^\dagger ({\bf r},t) \hat{\psi}_F ({\bf r},t) \rangle$ is
fermion density. Clearly, $\phi_B$ satisfies the Gross-Pitaevskii
equation
\begin{equation}
\left(\hat{H}_B^{(0)}+gn_B+fn_F \right) \phi_B=0 ,\label{gpe}
\end{equation}
where $n_B=|\phi_B|^2$.

Inserting Eqs.~(\ref{decom}) into (\ref{h}), neglecting terms
involving three or more fluctuation operators yields for
$\hat{\xi}_B({\bf r},t)$ and $\hat{\psi}_F ({\bf r},t)$ the
Heisenberg equations of motion:
\begin{eqnarray}
i\hbar \frac{\partial \hat{\xi}_B}{\partial t} = (\hat{H}_B-\mu_B
 ) \hat{\xi}_B + g\phi_B^2 \hat{\xi}_B^\dagger +f\phi_B
\delta \hat{\rho}_F ,\label{b} \\
i\hbar \frac{\partial \hat{\psi}_F}{\partial t} = ( \hat{H}_F-\mu_F
) \hat{\psi}_F + f(\phi_B \hat{\xi}_B^\dagger + \phi_B^* \hat{\xi}_B)
\hat{\psi}_F ,\label{f} \end{eqnarray}
with $\hat{H}_B=\hat{H}_B^{(0)}+2gn_B+fn_F$ and
$\hat{H}_F=\hat{H}_F^{(0)}+fn_B$.
 
We first find an approximate solution of (\ref{f}). Expanding
$\hat{\psi}_F ({\bf r},t)$ on the basis of eigenstates
$\varphi_n({\bf r})$, with eigenenergies $E_n$, of the Hamiltonian
$\hat{H}_F$ as:
\begin{equation}
\hat{\psi}_F ({\bf r},t)=\sum_n\,\hat{a}_n(t) \varphi_n({\bf r}),
\label{an}
\end{equation}
and inserting this expansion into Eq. (\ref{f}) yields the formal
solution
\begin{eqnarray}
\hat{\psi}_F ({\bf r},t)&=& \hat{\psi}_F^{(0)}({\bf r},t)-i\frac{f}{\hbar}
\int_0^t dt'\int d^3r'\,G({\bf r},{\bf r'},t-t') \nonumber \\
&&\;\; \times \,\Xi ({\bf
r'},t') \hat{\psi}_F({\bf
r'},t') , \label{psi}
\end{eqnarray}
where
\begin{eqnarray*}
& &\hat{\psi}_F^{(0)}({\bf r},t) = \sum_n \hat{a}_n(0) e^{-iE_n
t/\hbar} \varphi_n({\bf r}) ,\\ & & G({\bf r},{\bf r'},t-t') \equiv
\sum_n e^{-iE_n (t-t')/\hbar} \varphi_n({\bf r})\varphi_n^*({\bf
r'}), \\
& &\Xi ({\bf r'},t') = \phi_B({\bf r'})
\hat{\xi}_B^\dag({\bf r'},t') + \phi_B^*({\bf r'}) \hat{\xi}_B
({\bf r'},t').
\end{eqnarray*}
In first order, we replace $\hat{\psi}_F({\bf r'},t')$ by 
$\hat{\psi}_F^{(0)} ({\bf r'},t')$ on the r.h.s. of
Eq.~(\ref{psi}), so that
\begin{eqnarray}
\hat{\psi}_F ({\bf r},t) &\approx& \hat{\psi}_F^{(0)}({\bf
r},t)-i\frac{f}{\hbar} \int_0^t dt'\int d^3r'\,G({\bf r},{\bf r'},t-t')
\nonumber \\ &&\;\;\times \,\Xi ({\bf r'},t') 
\hat{\psi}_F^{(0)}({\bf r'},t') . \label{psi1}
\end{eqnarray}
After some algebra, this expression yields the lowest order
correction to the fermionic density in the presence of the Bose
condensate as:
\begin{eqnarray}
\delta \hat{\rho}_F &=& \hat{\psi}_F^\dagger ({\bf r},t) \hat{\psi}_F ({\bf
r},t)- n_F({\bf r}) \nonumber \\
&\approx& i\frac{f}{\hbar} \int_0^t dt' \int d^3r'\, {\cal J} ({\bf r},{\bf
r'},t-t') \Xi ({\bf r'},t') ,
\label{deltaF}
\end{eqnarray}
where
\begin{eqnarray*}
{\cal J}({\bf r},{\bf r'},t-t') &\equiv& G^*_>({\bf r},{\bf r'},t-t') G_< ({\bf
r},{\bf r'},t-t') \\
&&\;-G_>({\bf r},{\bf r'},t-t') G_<^*({\bf r},{\bf r'},t-t') ,\\
G_>({\bf r},{\bf r'},t-t') &\equiv& \sum_{\{ n|E_n > \mu_F \}}
e^{-iE_n(t-t')/\hbar}  \varphi^*_n({\bf r'}) \varphi_n({\bf r}),
\end{eqnarray*}
and $G_<$ is the same as $G_>$, but for $E_n \le \mu_F$. In deriving
Eq. (\ref{deltaF}), we have neglected second or higher order terms
in $\hat{\xi}_B$ and $\hat{\xi}_B^\dagger$, and we have replaced terms
of the form $\hat{\psi}_F^{(0)\dag} \hat{\psi}_F^{(0)}$ by their
expectation values with respect to the ground state. For example:
\begin{eqnarray*}
\hat{\psi}_F^{(0)\dag}({\bf r},t)\hat{\psi}_F^{(0)}({\bf r},t)
&\approx& \langle \hat{\psi}_F^{(0)\dag}({\bf
r},t)\hat{\psi}_F^{(0)}({\bf r},t) \rangle  = n_F ({\bf r}).
\end{eqnarray*}
Inserting Eq. (\ref{deltaF}) back into Eq.~(\ref{b}) yields the
equation of motion for the boson fluctuation operator:
\begin{eqnarray}
i\hbar \frac{\partial \hat{\xi}_B}{\partial t} &&= \left(\hat{H}_B^{(0)} +
2gn_B + fn_F \right)\hat{\xi}_B + g\phi_B^2 \hat{\xi}_B^\dagger \nonumber \\ &&
+ i\frac{f^2 }{\hbar} \int_0^t dt' \int d^3r'\, {\cal J} ({\bf r},{\bf
r'},t-t')  \Xi ({\bf r'},t')  \phi_B({\bf r}). \label{bb}
\end{eqnarray}

This is one of the central results of this paper. It shows that
the twofold effects of fermions on the excitation of the
condensate: First, the fermions change the effective mean-field
potential of the boson---the $fn_F$ term; second, the boson
fluctuations affect the fermions, which then results in a
back-action onto the bosons---the last term at the r.h.s. of
(\ref{bb}).

Consider for concreteness a homogeneous mixture of $N_B$ bosons
and $N_F$ fermions confined in a box of volume $V$. In the ground
state, we have $\mu_B = g n_B + f n_F$ and $ \mu_F = \hbar^2
k_F^2/(2m_F) + fn_B$, where we have introduced the 
densities $n_B = N_B/V$ and $n_F = N_F/V=k_F^3/(6\pi^2)$ and $k_F$
is the Fermi wave number. For periodic boundary conditions, the
eigenstates of $\hat{H}_F$ are plane waves $\varphi_{{\bf k}}
({\bf r}) =  e^{i {\bf k} \cdot {\bf r}} /\sqrt{V}$, with
eigenenergies $E_k = \hbar^2 k^2/2m_F + fn_B $ and $k=|{\bf k}|$, so that the
function ${\cal J}({\bf r},{\bf r'},t-t')$ of Eq.~(\ref{bb}) can now be
explicitly expressed as:
\begin{eqnarray*} 
{\cal J}({\bf r},{\bf
r'},t-t')=\frac{1}{V^2} &&\sum_{k_n>k_F} \sum_{k_m \le k_F} \left[
e^{i\frac{\hbar}{2m_F} (k_n^2-k_m^2)(t-t')}\right. \\
&& \left. e^{i({\bf k_m}-{\bf k_n}) \cdot {\bf r}} 
 e^{-i({\bf k_m}-{\bf k_n}) \cdot {\bf r'}} -c.c. \right] .
\end{eqnarray*}

Expanding likewise $\hat{\xi}_B ({\bf r},t)$ as:
\[ \hat{\xi}_B ({\bf r},t)=\frac{1}{\sqrt{V}} \sum_{{\bf k}} \eta_{{\bf k}}(t)
e^{i {\bf k}\cdot {\bf r}} ,\] yields from Eq. (\ref{bb})
\begin{equation}
i\hbar \frac{\partial \eta_{{\bf k}}}{\partial t} = {\cal L}_B
(k)\eta_{{\bf k}}+ gn_B \eta_{-{\bf k}}^\dag +(i/\hbar)f^2 n_B
I({\bf k}) ,\label{eta2}
\end{equation}
where ${\cal L}_B (k)=\hbar^2 k^2/2m_B+gn_B$,
\begin{eqnarray*}
I({\bf k}) &=& \frac{1}{V}\int_0^t dt' \int d^3r' \int d^3r \,e^{-i {\bf
k}\cdot {\bf r}} {\cal J}({\bf r},{\bf r'},t-t')  \\
&&\;\;\times \,\sum_{{\bf k'}} e^{i {\bf
k'}\cdot {\bf r'}} \left[ \eta^\dagger_{-{\bf k'}} (t') + \eta_{{\bf k'}}(t')
\right] \\
&=&\frac{1}{V}\int_0^t dt'\, \sum_{ k_m \le k_F}
\left[e^{i\frac{\hbar}{2m_F}  (|{\bf k_m} + {\bf k}|^2-k_m^2)
(t-t')}-c.c. \right]  \\
&&\;\;\times \, \Theta (|{\bf k_m} + {\bf k}|-k_F)  
\left[ \eta^\dagger_{-{\bf k}} (t') + \eta_{{\bf k}}(t') \right]  ,
\end{eqnarray*}
and $\Theta(x)$ is the unit step function. Similarly,
\begin{equation}
i\hbar \frac{\partial \eta^\dag_{-{\bf k}}}{\partial t} = -{\cal
L}_B (k)\eta^\dag_{-{\bf k}}- gn_B \eta_{{\bf k}} -(i/\hbar)f^2
n_B I({\bf k}) .\label{eta2+}
\end{equation}

The two coupled integro-differential equations (\ref{eta2}) and
(\ref{eta2+}) can be solved using Laplace transformation. Denoting
$
\alpha_{{\bf k}}(s) = L[\eta_{{\bf k}}(t)] =\int^{\infty}_0
dt\,e^{-st} \eta_{{\bf k}}(t)$ and $\beta_{{\bf k}}(s) =
L[\eta^\dagger_{-{\bf k}}(t)] $, Eqs.~(\ref{eta2}) and (\ref{eta2+})
yield
\begin{mathletters}
\begin{eqnarray}
i\hbar s \alpha_{{\bf k}}(s) &=& {\cal L}_B(k) \alpha_{{\bf k}}(s) +gn_B
\beta_{{\bf k}}(s)  \nonumber \\
&&\;\; +i\frac{f^2}{\hbar}n_B \ell_s(k)[\alpha_{{\bf k}}(s)+
\beta_{{\bf k}}(s)] , \\
i\hbar s\beta_{{\bf k}}(s) &=& -{\cal L}_B(k)\beta_{{\bf k}}(s)  -gn_B
\alpha_{{\bf k}}(s) \nonumber \\
&&\;\; - i\frac{f^2}{\hbar}n_B \ell_s(k) [\alpha_{{\bf
k}}(s)+ \beta_{{\bf k}}(s)],
\end{eqnarray}
\label{s}
\end{mathletters}
where
\begin{eqnarray*}
\ell_s(k) = \frac{1}{V} L\left[\sum_{k_m \le k_F}^{|{\bf
k_m}+{\bf k}| > k_F}
e^{i\Delta_{{\bf k}+{\bf k_m},{\bf k_m}}t} 
-c.c. \right], \end{eqnarray*}
where $\hbar \Delta_{{\bf k}+{\bf k_m},{\bf k_m}}= \epsilon_{{\bf 
k_m}+{\bf k}}-\epsilon_{{\bf k_m}}$ with $\epsilon_{{\bf k}} = 
\hbar^2 k^2/2m_F$.
The excitation frequency $\omega$ is simply obtained by replacing
$s$ by $i\omega + 0^+$ in Eqs.~(\ref{s}). Since these equations
are homogeneous equations in $ \alpha_{{\bf k}}(s)$ and
$\beta_{{\bf k}}(s) $, $\omega$ is determined by requiring the
determinant of their coefficients to be equal to zero. These steps
result in the phonon dispersion relation:
\begin{equation}
\hbar^2\omega^2=\frac{\hbar^2
k^2}{2m_B} \left[\frac{\hbar^2 k^2}{2m_B}+ 2gn_B+\frac{f^2 n_B}{(2\pi)^3
\hbar} \ell_{\omega}(k) \right] ,\label{o1}
\end{equation}
where
\begin{eqnarray}
\ell_{\omega}(k) &=& \int d{\bf k_m} \,\left( \frac{\Theta(|{\bf k_m}+{\bf
k}|-k_F)} {\omega-\Delta_{{\bf k}+{\bf k_m},{\bf k_m}}-i0^+} \right.\nonumber
\\ &&\,- \left.
\frac{\Theta(|{\bf k_m}+{\bf k}|-k_F)} {\omega+\Delta_{{\bf k}+{\bf
k_m},{\bf k_m}}-i0^+}  \right) \Theta(k_F - k_m) .
\label{lok}
\end{eqnarray}
Here we have replaced the sum over ${\bf k_m}$ by
an integral. As expected the Bogoliubov spectrum of a pure
condensate is recovered for $f=0$.

Equation (\ref{lok}) has the same form as Eq.~(12.30) in Ref.~\cite{fetter}. 
In general, $\ell_{\omega}(k)$---and hence $\omega$---may take
complex values, indicating that the homogeneous state of the Bose-Fermi
mixture becomes unstable. Its real part is given by the principal
value of the integral in (\ref{lok}) and can be expressed as:
\begin{eqnarray*}
{\rm Re} &&[\ell_{\omega}(k)] =
-\frac{4 \pi m_F
k_F}{\hbar}-\frac{2\pi}{\alpha^3}\left[ \left(\omega-\frac{\alpha
k}{2}\right)^2-\alpha^2 k_F^2 \right] \\
&& \times \,\ln \left|
\frac{\omega+\alpha(k_F-k/2)}{\omega-\alpha(k_F+k/2)} \right|
-\frac{2\pi}{\alpha^3} \left[ \left(\omega+\frac{\alpha
k}{2}\right)^2-\alpha^2 k_F^2 \right] \\
&& \times \,\ln
\left| \frac{\omega-\alpha(k_F-k/2)}{\omega+\alpha(k_F+k/2)} \right| ,
\end{eqnarray*}   
with $\alpha=\hbar k/m_F$. Its imaginary part, which determines
the stability of the mixture, is given by
\begin{eqnarray}
{\rm Im} [\ell_{\omega}(k)] &= & \pi \int d{\bf k_m}  \,\Theta(|{\bf k_m}+{\bf
k}|-k_F) \Theta(k_F-k_m) \nonumber \\
& & \times \,\delta\left( \omega- \Delta_{{\bf k}+{\bf k_m},{\bf
k_m}} \right),
 \label{im}
\end{eqnarray}
which is nothing but the dynamical structure factor of the Fermi gas,
$S(k,\omega)$, apart from a constant factor\cite{pines}.
\begin{figure}
\begin{center}
    \includegraphics*[width=0.72\columnwidth,
height=0.4\columnwidth]{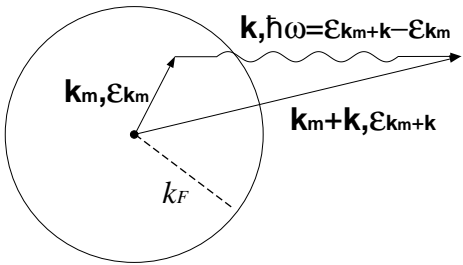}
\vspace{3 mm}
\caption{Resonant phonon-exchange interactions between fermions results in
dynamical instability of the system.} \label{fig0}
\end{center}
\end{figure}
The physical interpretation of the instability associated with
${\rm Im} [\ell_{\omega}(k)]\neq 0$ [or $S(k,\omega) \neq 0$] is directly
apparent from  Eq.~(\ref{im}). The condition that a phonon of wave vector
${\bf k}$ and frequency $\omega$ satisfies $ \omega= \Delta_{{\bf k}+{\bf
k_m},{\bf k_m}}$ for some $k_m < k_F$ and $|{\bf k_m}+{\bf k}|> k_F$
implies that it resonantly couples a pair of fermions lying inside
and above the Fermi sea, respectively, see Fig.~\ref{fig0}. 
When
this coupling occurs, the Bose-Fermi mixture becomes dynamically
unstable, and the system will spontaneously break into several
distinct phases\cite{smith}. For long-wavelength phonons ($k \ll
k_F$), the stability criterion determined by ${\rm Im}
[\ell_{\omega}(k)] =0$ requires the phonon frequency to satisfy:
\begin{equation} \omega > \frac{\hbar k^2}{2m_F} +
\frac{\hbar k k_F}{m_F}  .
\label{cri} \end{equation}
\begin{figure}
\begin{center}
    \includegraphics*[width=0.95\columnwidth,
height=0.6\columnwidth]{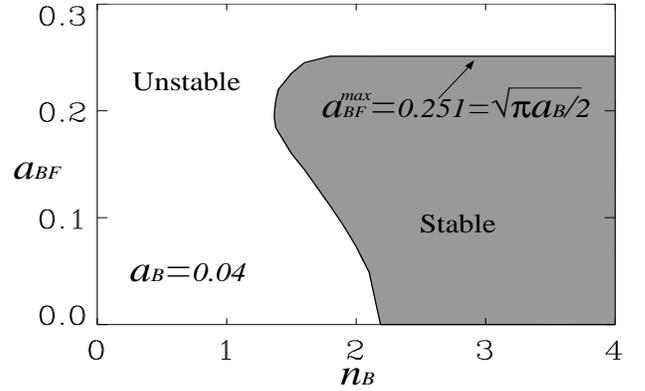}
\vspace{3 mm}
\caption{Stability diagram of a Bose-Fermi mixture. We assume $m_F=m_B=m$. We
have adopted a system units in which the units for frequency, length, and
wavenumber are $\hbar k_F^2/2m$, $1/k_F$, and $k_F$, respectively. Note that
in this units, the fermion density is given by $n_F=1/(6\pi^2) \approx
0.017$.}
\label{fig1}
\end{center}
\end{figure}
The stability regime and the phonon spectrum can be obtained by
first solving Eq. (\ref{o1}) numerically while assuming ${\rm Im}
[\ell_{\omega}(k)]=0$, and then checking if the system is stable
using the criterion (\ref{cri}). Fig.~\ref{fig1} shows the
stability diagram of the mixture in the $a_{BF}-n_B$ space. As can
be seen, the dynamical stability of the system is determined by
both the scattering lengths and the densities. All other
parameters being fixed, the stability condition (\ref{cri})
imposes a minimum boson density $n_B^{min}$ beyond which no stable
homogeneous mixture exists. We can use the Bogoliubov spectrum of
a pure condensate to estimate $n_B^{min}$ as   \[g
n_B^{min}\approx \frac{\hbar^2 k_F^2}{m_F^2/m_B} =\frac{\hbar^2
(6\pi^2 n_F)^{2/3}}{m_F^2/m_B}.\] For realistic numbers,
$n_B^{min}$ is about two orders of magnitude larger than $n_F$
(see Fig.~\ref{fig1}).

\begin{figure}
\begin{center}
    \includegraphics*[width=0.95\columnwidth,
height=0.6\columnwidth]{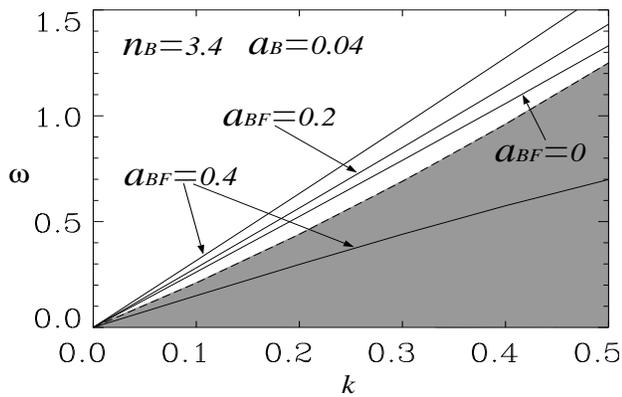}
\vspace{3 mm}
\caption{Phonon spectrum of a boson-fermion mixture. The dashed line
corresponds to $\omega=\hbar(k^2+2kk_F)/2m_F$. The gray area below the dashed
line represents the unstable regime. Same units as in Fig.~\ref{fig1}.}
\label{fig2}
\end{center}
\end{figure}
Figure~\ref{fig2} illustrates the phonon spectrum in the
Bose-Fermi mixture for $n_B>n_B^{min}$. For small values of
$a_{BF}$, the spectrum is stable and single valued.  Furthermore,
the boson-fermion interaction increases the sound velocity of the
phonon. By expanding $\ell_{\omega}(k)$ around small $k$, it is
not difficult to show that in our dimensionless units (see caption
of Fig.~\ref{fig1}) and for $m_B=m_F$  the sound velocity is
approximately given by
\begin{equation}
c \approx c_0 \left[1+\frac{4a_{BF}^2}{(2+c_0)^2\pi a_B}
\right],
\label{sound}
\end{equation}
where $c_0=\sqrt{16\pi a_B n_B}$ is the sound velocity in a pure
condensate.  For the parameters in Fig.~\ref{fig2}, we have a 6\%
increase in sound velocity if $a_{BF}$ changes from 0 to 0.2.
However, further increasing $a_{BF}$ beyond a critical value
$a_{BF}^{max}$ splits the phonon spectrum into two branches, and
one of them falls into the unstable regime. Again using the
expansion of $\ell_{\omega}(k)$, one can show that $a_{BF}^{max}$
is determined by (in dimensionless units):
\[(a_{BF}^{max})^2 = 2\pi a_B m_r/(m_B+m_F) .\]
This condition agrees with that obtained by Viverit {\it et al}.\cite{smith}
and is confirmed by our numerical calculations (see Fig.~\ref{fig2}). For $m_F
\approx m_B$, we have \begin{equation}
a_{BF}^{max}=\sqrt{\pi a_B/2}.
\label{amax}
\end{equation}
Note that while $n_B^{min}$ is independent of $a_{BF}$, $a_{BF}^{max}$ is
independent of $n_B$. Putting (\ref{amax}) into (\ref{sound}), we obtain the
maximum sound velocity achievable in a homogeneous mixture as:
\[ c_{max} = c_0 \left[ 1+ \frac{2}{(2+c_0)^2} \right] .\]

In conclusion, we have developed a general formalism to determine
the phonon spectrum of the condensate in a Bose-Fermi mixture at
zero temperature.  We have shown that a resonant phonon-exchange
interaction between fermions can induce a dynamical instability in
the mixture. As an example, we have calculated the phonon spectrum
of a homogeneous system and obtained analytically the sound
velocity under the effect of boson-fermion coupling. From the
spectrum, we are able to determine the dynamical stability regime
of such a system which depends on both the scattering lengths and
the densities, and is directly linked to the dynamical structure factor of the
Fermi gas. Our calculations show that for a given fermion density, a much
denser boson field is required to stabilize the mixture. Viverit {\em et al.}
also considered the stability of a homogeneous mixture\cite{smith}. In
contrast to our work, which depicts a full dynamic picture for the
boson-fermion mixture, their interest was focused on the mechanical stability
phase diagram of the system, which is obtained by requiring the total energy
to be a minimum but does not take the phonon-exchange interaction into full
account.

Our method can
be straightforwardly applied to an inhomogeneous system, although the algebra
in this case is much more involved. We will address this situation in a
future publication. Another interesting case is when a boson field is mixed
with a fermion field containing two components (e.g., two spin states).
Interactions between the bosons and fermions will induce an
effective coupling between the fermions from different components,
which may result in the creation of Cooper pairs.

This work is supported in part by the US Office of Naval Research under
Contract No. 14-91-J1205, by the National Science Foundation under Grant
No. PHY98-01099, by the US Army Research Office, and by the Joint Services
Opitcs Program.One of us (MW) acknowledges
support by the DFG national programme SPP 1116.


\end{document}